\documentclass[fleqn,usenatbib]{mnras}

\usepackage{newtxtext,newtxmath,subfig}

\usepackage[T1]{fontenc}
\usepackage{ae,aecompl}

\usepackage{graphicx}	
\usepackage[utf8]{inputenc} 
\usepackage{appendix}

\title{On Modelling Cosmic Ray Sputtering of Interstellar Grain Ices}

\author[A. Paulive, J. T. Carder, E. Herbst]{
Alec Paulive,$^{1}$\thanks{E-mail:ap4kz@virginia.edu}
{Joshua T. Carder,$^{1}$}
{Eric Herbst,$^{1,2}$}
\\
$^{1}$Department of Chemistry, University of Virginia, McCormick Road, Charlottesville, VA 22904, USA \\
$^{2}$ Department of Astronomy, University of Virginia, McCormick Road, Charlottesville, VA 22904, USA \\
}

\date{Accepted XXX. Received YYY; in original form ZZZ}

\pubyear{2022}

\begin{document}
\label{firstpage}
\pagerange{\pageref{firstpage}--\pageref{lastpage}}
\maketitle

\begin{abstract}
In the interstellar medium (ISM), the formation of complex organic molecules (COMs) is largely facilitated by surface reactions. However, in cold dark clouds, thermal desorption of COMs is inefficient because of the lack of thermal energy to overcome binding energies to the grain surface. Non-thermal desorption methods are therefore important explanations for the gas-phase detection of many COMs that are primarily formed on grains. Here we present a new non-thermal desorption process: cosmic ray sputtering of grain ice surfaces based on water, carbon dioxide, and a simple mixed ice. Our model applies estimated rates of sputtering to the 3-phase rate equation model \texttt{Nautilus-1.1}, where this inclusion results in enhanced gas phase abundances for molecules produced by grain reactions such as methanol (CH$_{3}$OH) and methyl formate (HCOOCH$_{3}$). Notably, species with efficient gas phase destruction pathways exhibit less of an increase in models with sputtering compared to other molecules. These model results suggest that sputtering is an efficient, non-specific method of non-thermal desorption that should be considered as an important factor in future chemical models. 

\end{abstract}

\begin{keywords}
ISM: molecules -- astrochemistry -- ISM: cosmic rays
\end{keywords}



\section{Introduction}

In cold dark clouds, recent detections of complex organic molecules including some C$_{2}$H$_{4}$O$_{2}$ isomers have led to an influx of studies examining chemical formation routes \citep{jimenez-serra_spatial_2016, soma_complex_2018}. Warmer sources have well-understood thermal production routes for species formed both in the gas phase and on grains, with efficient grain processes producing COMs, which then thermally desorb with the majority of ice from dust grains \citep{charnley_molecular_1992}. The production of the same COMs in cold dark clouds has been studied, yet continues to be an area of interest, if not fully understood \citep{vasyunin_reactive_2013,chang_unified_2016,balucani_formation_2015, Jin_Garrod_2020}. Desorption from grain mantles is a common problem in dark clouds, as most models still under-produce gas-phase abundances of COMs, though non-thermal desorption methods are consistently being developed and applied to models of cold dark clouds, with a particular emphasis on reactive desorption and photodesorption \citep{garrod_non-thermal_2007,oberg_photodesorption_2009}.  
 
Cosmic rays are highly energetic ions travelling at significant portions of the speed of light, with H$^{+}$ being the most common ion \citep{cummings_galactic_2016}. Their role in interstellar grain chemistry has been increasingly examined, with radiolysis theory and experiments highlighting the importance of cosmic ray and grain interactions in generating chemical diversity through ionization and excitation of ices \citep{hudson_ir_2005, abplanalp_formation_2015, boyer_role_2016, shingledecker_general_2018}. When cosmic rays collide with interstellar ices and dust grains, perhaps the most noticeable effect is the heating of the aggregate, referred to as whole grain cosmic ray heating or simply cosmic ray heating (CRH), causing periods of enhanced thermal activity, such as desorption and possibly enhanced thermal diffusion \citep{hasegawa_new_1993, kalvans_temperature_2018}.

 
The energy deposited to the surface by cosmic rays does not just cause thermal heating, ionization, and excitation; sputtering is a direct outcome of cosmic ray interactions with grain ices. Sputtering is defined as the desorption of material resulting from high energy particle impacts. This is a separate process from the thermal desorption resulting from the heating of the grains, as the act of collision itself causes not just heating, but also the ejection of surface molecules. Multiple experiments have been recently done to examine the sputtering of high-energy cosmic rays ($\ge$ MeV) impacting interstellar ices, and have shown that sputtering is an efficient method of desorption from surfaces \citep{dartois_swift_2013,dartois_heavy_2015}. Even more recently, chemical models including experimental sputtering data were utilized, demonstrating increased gas-phase abundances of various molecules in chemical kinetic models, although the experiments used heavier ions than the most common cosmic ray, H$^+$ \citep{dartois_cosmic_2018, wakelam_efficiency_2021}.
 
The many theories of sputtering predict that the interaction of cosmic rays and ice surfaces varies based on the energy and mass of the incoming particle, resulting in competition between sputtering caused by high energy ($\ge$ MeV) particles and less energetic particles in the keV range. The competing regimes are further divided by two mechanisms of energy transfer: electronic sputtering, where ineleastic collisions cause movement from coulombic interactions, and nuclear sputtering, where elastic collisions between nuclei cause physical recoil and desorption \citep{sigmund_theory_1969, brown_sputtering_1978}. Lower energy particles more efficiently deposit energy into surfaces through elastic (nuclear) collisions, while for most cosmic ray energies, inelastic (electronic) collisions are the main source of energy transfer \citep{andersen_sputtering_1981, johnson_energetic_1990}. Reasons for the discrepancy in effectiveness will be examined in a later section. 
 

In this paper, we examine sputtering theory by including sputtering into a rate-equation based 3-phase model under cold dark cloud conditions. Section ~\ref{sec:model} presents the theoretical treatment of sputtering used in the model along with all physical parameters and chemical networks utilized.
Section ~\ref{sec:results} covers the results of the models for theories based on water, carbon dioxide, and a theoretical mixed ice.
Section ~\ref{sec:analysis} discusses the implications sputtering has on interstellar environments with comparisons to astronomical observations, and Section ~\ref{sec:conclusions} concludes our work.


\section{Models}
\label{sec:model}

The chemical models reported in this paper make use of the three-phase rate equation-based gas-grain program \texttt{Nautilus-1.1} \citep{wakelam_kinetic_2012,ruaud_gas_2016}. The three phases comprise the gas, the top layers of the grain ice (two monolayers in this instance) and the remaining ice on the dust grain mantle, called the bulk. This distinction is relevant to diffusion of ice species in the mantle, and a variety of adsorption and desorption methods utilized in \texttt{Nautilus-1.1}. These methods include photodesorption, reactive desorption, and thermal desorption, which allow for exchange of species between the ices upon the grain surfaces and the gaseous interstellar medium, are further described later in the section \citep{vasyunin_reactive_2013}. Sputtering, which removes particles from the ice surface and bulk, is slightly different from already included non-thermal desorption methods due to the frequency of sputtering events, and the number of species removed from both the ice surface and bulk. These will be examined further in Section~\ref{sec:sputtering}. 

With the Nautilus code, we are able to run a series of chemical models with specific parameters in which the diverse effects of cosmic ray interactions in various combinations of activity or inactivity on sputtering are explored. Further details, which apply to all models, such as initial abundances, and constant physical conditions are shown in Table~\ref{tab:abundances} and \ref{tab:physicalcond}. These physical conditions, which are the same as in \citet{ruaud_gas_2016}, are comparable to those observed in cold, dark molecular clouds, which are the environments on which we focus.

\begin{table}
    \centering
    \caption{Initial abundances of elements with respect to total hydrogen nuclei}
    \begin{tabular}{lcr}
        \hline
         Species & Abundance \\
        \hline
         H$_{2}$ & $0.499$ $^{\textup{a}}$ \\
         He & $9.000 \times 10^{-2} $ $ {^\textup{a}}$ \\
         N & $6.200 \times 10^{-5} $ $ {^{\textup{a}}} $ \\
         C$^{+}$ & $1.700 \times 10^{-4} $ $^{\textup{b}} $ \\
         O & $2.429 \times 10^{-4} $ $ ^{\textup{c}} $ \\
         S$^{+}$ & $8.000 \times 10^{-8}$ $^{\textup{d}} $ \\
         Na$^{+}$ & $2.000 \times 10^{-9}$ $^{\textup{d}} $ \\
         Mg$^{+}$ & $7.000 \times 10^{-9}$ $^{\textup{d}} $ \\
         Si$^{+}$ & $8.000 \times 10^{-9}$ $^{\textup{d}} $ \\
         P$^{+}$ & $2.000 \times 10^{-10}$ $^{\textup{d}} $ \\
         Cl$^{+}$ & $1.000 \times 10^{-9}$ $^{\textup{d}} $ \\
         Fe$^{+}$ & $3.000 \times 10^{-9}$ $^{\textup{d}} $ \\
         F & $6.680 \times 10^{-9}$ $^{\textup{e}} $ \\
        \hline
        $^{\textup{a}}$ \citep{wakelam_polycyclic_2008} \\
        $^{\textup{b}}$ \citep{jenkins_unified_2009} \\
        $^{\textup{c}}$  \citep{hincelin_oxygen_2011}  \\
        $^{\textup{d}}$ \citep{graedel_kinetic_1982} \\
        $^{\textup{e}}$ \citep{neufeld_chemistry_2005} \\
    \end{tabular}
    \label{tab:abundances}
\end{table}

\begin{table}
    \centering
    \caption{Physical conditions utilized in models for this work, based on TMC-1 conditions}
    \begin{tabular}{lccr}
        \hline
         Parameter & TMC-1 \\
        \hline
         $\textit{n}_{\textup{H}}$ (cm$^{-3}$) & $10^{4}$ \\
         $\textit{n}_{\textup{dust}}$ (cm$^{-3}$) & $1.8 \times 10^{-8}$ \\
         $\textit{T}_{\textup{gas}}$ (K) & $10$ \\
         $\textit{T}_{\textup{dust}}$ (K) & $10$ \\
         $\textit{N}_\textup{{site}}$ (cm$^{-2}$) & $1.5 \times 10^{15}$ \\
         $\zeta$ (s$^{-1}$) & $1.3 \times 10^{-17}$ \\
         $\textit{A}_{\textup{V}}$ & 10 \\
        \hline
    \end{tabular}
    \label{tab:physicalcond}
\end{table}

To accurately gauge the impact of each new or updated process examined in this paper, we have devised separate models with differing processes. These are shown in Table~\ref{tab:model_key}, which lists the name by which the specific model will be referenced in the paper. The table also indicates the inclusion or absence of cosmic ray whole grain heating (CRH) and sputtering in the labelled models \citep{hasegawa_new_1993}. Our models include some features not in the standard version of \texttt{Nautilus-1.1}. These additional options include a competitive tunneling mechanism between activation and diffusion barriers set to the faster option, and an option that prevents all species but H and H$_2$ from tunneling under activation energy barriers of surface reactions, set to "on" \citep{smith_chemistry_2008,shingledecker_simulating_2019}. Nonthermal desorption mechanisms for surface species also have associated options. Photodesorption \citep{bertin_indirect_2013} and cosmic ray induced photodesorption \citep{hasegawa_new_1993} are included, along with reactive desorption at 1 per cent probability \citep{garrod_non-thermal_2007}. These three switches are set to "on", with both the photodesorption yield, and the cosmic ray induced photodesorption yield set at $1 \times 10^{-4}$. All models have the same parameters for the features described above, except in the cases of sputtering and CRH.

\begin{table}
    \centering
    \caption{Model identifiers with a listing of the active sputtering rate and whether cosmic ray heating was included}
    \begin{tabular}{lcc}
       \hline
        Model & Sputtering & Note \\
       \hline
        1H & Off & Base Model \\
        2HS & On - H$_{2}$O Ice & CRH \\
        3S & On - H$_{2}$O Ice & No CRH \\
        4HSC & On - CO$_{2}$ Ice & CRH \\
        5SC & On - CO$_{2}$ Ice &  No CRH \\
        6HSM & On - Mixed Ice & CRH \\
        7SM & On - Mixed Ice & No CRH \\
        \hline
    \end{tabular}
    \label{tab:model_key}
\end{table}

\subsection{Network}
\label{sec:network}

The \texttt{Nautilus-1.1} reaction network used in this study has been expanded to include more complex chemical species that lead up to the C$_{2}$H$_{4}$O$_{2}$ isomers, methyl formate (HCOOCH$_{3}$), glycolaldehyde (HCOCH$_{2}$OH), and acetic acid (CH$_{3}$COOH), as described in \citet{paulive_role_2021}. These molecules are among a class of species known as complex organic molecules (COMs), which range from 6-13 atoms and are partially saturated.The base network of gaseous reactions is taken from the KIDA network \citep{wakelam_kinetic_2012}. The granular reaction network is from the \texttt{Nautilus-1.1} package, with additional thermal grain-surface reaction pathways leading to C$_{2}$H$_{4}$O$_{2}$ isomer precursors from \citet{garrod_formation_2006}. Used initially in hot core models, these reactions have been included here to provide likely thermal pathways to COMs, which may eventually leave the grain surface through cosmic ray interactions \citep{skouteris_genealogical_2018}. Species with a prefix of ``J'' are those on the surface of the ice mantle, while species with a ``K'' lie within the bulk of the ice, which, after the simulation has completed running ($10^7$ yrs), contains close to 100 monolayers of ice. Desorption energies, $E_{\rm D}$, are from a combination of \citet{garrod_formation_2006,garrod_complex_2008,garrod_three-phase_2013}. The diffusion barriers ($E_{\rm b}$) are $0.4 \times E_{\rm D}$ for ice surface species and $0.8 \times E_{\rm D}$ for bulk species. All ices in this paper are considered amorphous solids.

\subsection{Sputtering Theory}
\label{sec:sputtering}

As noted above, sputtering is the process of impacting a surface (or surface of a liquid) with high energy particles resulting in the ejection of particles from the surface into the surrounding space. By applying sputtering to grain surfaces in the interstellar medium, it is possible to estimate rate coefficients, used to calculate rates, based on cosmic ray grain interaction rates, stopping powers, and average energy loss due to collisions, all of which can be used in rate-equation based models. In these models, the overall rate of desorption (R$^{\textup{tot}}$) for some mantle species, $i$, is the sum of all desorption processes. Example desorption mechanisms include thermal desorption (R$^{\textup{Th}}$, photodesorption (R$^\textup{{Pho}}$), and reactive desorption (R$^\textup{{RD}}$). With the addition of sputtering in our models, the calculation of the rate of sputtering (R$^\textup{{s}}$) is necessary. The overall rate of desorption for species $i$ is as follows: 

\begin{equation}
    \mathrm{ R^{tot} }= - \frac{dN_{i}}{dt} = \mathrm{ R^{Th} + R^{Pho} + R^{RD} ... + R^{s}}.
    \label{differential}
\end{equation}

\noindent Here $\frac{dN_i}{dt}$ is the rate of change of the gas-phase concentration of desorbing species $i$, with units of cm$^{-3}$ s$^{-1}$. The individual rate processes, R, are each given by first-order rate coefficients, $k$, multiplied by N$_{i}$, so that the total rate, R$^{\textup{tot}}$, can be written by 

\begin{equation}
    \mathrm{ R^{tot} }= - \frac{dN_{i}}{dt} = N_{i} \left( k_{Th} + k_{Pho} + k_{RD} ... + k_{s} \right).
    \label{differential2}
 \end{equation}

\noindent Here $k_{\textup{Th}}$, $k_{\textup{Pho}}$, $k_{\textup{RD}}$, and $k_{\textup{s}}$ are the rate coefficients for thermal desorption, photodesorption, reactive desorption, and sputtering, respectively. We will examine how to calculate $k_{\textup{s}}$ later in the section. It is important to note that our models assume all species are affected by sputtering at the same rate coefficient. When referring to the rate coefficient of sputtering off of water ice, carbon dioxide ice, or mixed ice, all ice mantle species in the model will take on the respective sputtering rate for the assumed ice composition. For example, when referring to water ice sputtering, the model implements this by setting the rate coefficient of all ice mantle species to be the calculated rate coefficient for water ice. Models with CO$_{2}$ ice sputtering set the rate coefficient of all species to be equal to that of CO$_{2}$ ice.

The units of concentration for species $i$ can vary, so long as they are consistent throughout the model. The number of molecules on a grain, the fractional abundance to hydrogen or water, and the relative density are all common. For this model the concentration has the units of cm$^{-3}$.

Some basis for the assumption that all ice species in the model take on the respective sputtering rate is given in the next paragraph. The assumption is justified by two factors: first is the physical structure of amorphous solids. \citet{clements_kinetic_2018} shows that the formation of interstellar ices at 10 K results in a highly porous structure, creating large "chunks" of ice that are interconnected to form the ice mantle. The ices themselves are observed to be approximately 70\% water ice \citep{1996A&A...315L.357W}, meaning that if there is a water ice molecule that is successfully removed from either the surface or the mantle, there is a probability that the water molecule will be removed with neighboring species, or if the ice is porous enough, a large "chunk" of ice will be desorbed alongside the sputtered molecule.
Second is experimental evidence regarding sputtering yields. There are multiple studies on the effectiveness of sputtering multiple molecules and dimers from a chemisorbed metal surface \citep{gades_dimer_1995}. Recent experiments with swift heavy ions colliding with amorphous ices and crystalline water ice also sputter large quantities of ice, about 20,000 water molecules per incident ion, although the high number of desorbed molecules is not attributed to the porosity of the ice \citep{dartois_swift_2013,dartois_swift_2015, dartois_cosmic_2018}.
These two factors suggest that sputtering has the possibility of desorbing multiple ice species per incident cosmic ray because water, the main component of the ice, will either be desorbed alongside whatever else it is next to, or in the same cluster as the water molecules that are sputtered.


We can approximate a first-order sputtering rate coefficient, $k_{s}$, in units of cosmic ray particles per second, starting with a cosmic ray flux for the interstellar medium $\phi_{ism}$ (cosmic ray particles s$^{-1}$ cm$^{-2}$) and a cross section for the cosmic ray interacting with a molecule on the grain surface $\sigma$ (cm$^{2}$): 

\begin{equation}
    k_{s} = \sigma \phi_{ism}.
    \label{eq:basic_kinetics}
\end{equation}

We obtain a value for the cosmic ray flux by integrating the Spitzer-Tomasko energy function $j(E)$ \citep{spitzer_heating_1968} over the cosmic ray energy distribution from 0.3 MeV ($3 \times 10^{-4}$ GeV) to 100 GeV given by the equation:

\begin{equation}
    \label{eq:cr_e_dist}
    j(E) = \frac{0.90}{\left(0.85+E_{G}\right)^{2.6}} \frac{1}{\left(1+0.01/E_{G}\right)}.
\end{equation}

\noindent Here $E_{\textup{G}}$ is the energy of cosmic rays in GeV. The distribution is graphed in Figure~\ref{fig:cr_dist}. Assuming that cosmic rays are isotropic and then integrating over the energy distribution results in an interstellar cosmic ray flux $\phi_{st}$ of 8.6 particles cm$^{-2}$ s$^{-1}$. An additional term, $\zeta$, based on the cosmic ray ionization rate of $\approx 1.36 \times 10^{-17}$ s$^{-1}$ allows for easy scaling of the cosmic ray flux depending on the ionization rate in varying areas of the ISM, similar to the scaling factor included in \citet{shingledecker_general_2018}, where the overall flux term is as follows:

\begin{figure}
    \centering
    \includegraphics[width=\columnwidth]{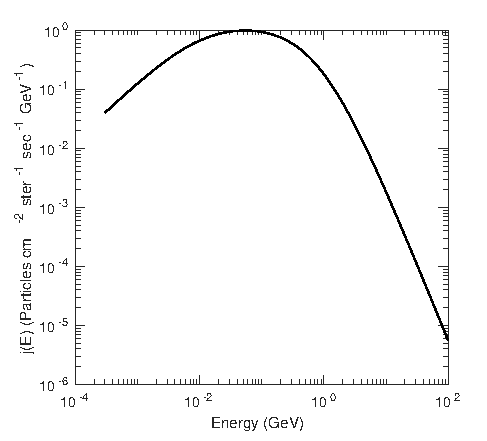}
    \caption{Cosmic ray energy distribution function, from \citet{spitzer_heating_1968}. The cosmic ray flux in particles cm$^{-2}$ s$^{-1}$ is obtained by integrating j(E) over the cosmic ray energy distribution, from $3 \times 10^{-4}$ GeV to 100 GeV. This results in an overall cosmic ray flux of 8.6 particles cm$^{-2}$ s$^{-1}$.}
    \label{fig:cr_dist}
\end{figure}

\begin{equation}
    \label{eq:scaling_factor}
    \phi_{ism} = \phi_{st} \frac{\zeta}{1.36 \times 10^{-17}} . 
\end{equation}

Obtaining an exact sputtering cross section that will work with our averaged model is difficult, as the sputtering cross section is dependent on the stopping power of the target ice, as well as the mass and energy of the incident cosmic ray \citep{townsend_comparisons_1993,andersen_sputtering_1981}. Although there are experimentally determined sputtering cross sections, they are difficult to incorporate into our model for a number of reasons.  First, \texttt{Nautilus-1.1} does not take into account specific cosmic ray nuclei. Secondly, many experimental and theoretical cross sections are not specific to sputtering, instead measuring all manners of destruction, called a destruction cross section. This results in "double counting" other included desorption processes and radiolysis effects if this total destruction cross section were to be used for calculating sputtering rates. 

A total cross section $\sigma_{t}$ is split into two general categories for collisions: a nuclear cross section $\sigma_{n}$, for elastic collisions, and an electronic (inelastic) cross section $\sigma_{e}$. The electronic cross section can be further split into ionizing $\sigma_{\textup{ion}}$ and excitation $\sigma_{\textup{exc}}$ cross sections.  The different partial cross sections are shown in the following equation:

\begin{equation}
    \label{eq:cross_sec}
    \sigma_{t} = \sigma_{n} + \sigma_{e} = \sigma_{n} + \sigma_{\textup{ion}} + \sigma_{\textup{exc}}.
\end{equation}

\noindent These cross sections all contribute to physical effects such as sputtering and whole grain heating. The fraction of each cross section that contributes to each resulting interaction type is not well defined.

In a similar manner to splitting the total cross section into more focused cross sections to describe different processes, it is helpful to divide $E_{\textup{total}}$, the total energy loss, into more specific terms:

\begin{equation}
    \label{eq:e_partition}
    E_{\textup{total}} = N_{\textup{ion}} \Bar{E}_{\textup{ion}} + N_{\textup{exc}} \Bar{E}_{\textup{exc}} + N_{\textup{ion}} \Bar{E}_{s} + \nu_{\textup{n}}(E_{\textup{total}})
\end{equation}

\noindent where $N_{\textup{ion}}$ and $N_{\textup{exc}}$ are the numbers of ionizations and excitations, respectively, while the average energies lost to ionization and excitation are $\Bar{E}_{\textup{ion}}$ and $\Bar{E}_{\textup{exc}}$. $\Bar{E_{s}}$ is the energy lost to sub-excitation secondary electrons, or free electrons resulting from ionization that are too low in energy to cause further ionizations or excitations. These secondary electrons eventually lose their energy to the surface as vibrational or rotational energy, and these directly contribute to the motions of species on the ice surface, which can result in sputtering. The sum of the contributions of all contributing electronic effects on a single molecule can be referred to as the average electronic energy loss per ion-pair, or $W_{\textup{e}}$, with units of eV. $W_{\textup{e}}$ for a single water molecule is approximately 30 eV \citep{johnson_electronic_1982}. 

The remaining term, $\nu_{\textup{n}}(E_{\textup{total}})$, where $\nu_{\textup{n}}$ is the fraction of $E_{\textup{total}}$ lost by nuclear effects, is the energy lost due to nuclear elastic collisions, and is related to the nuclear stopping power cross section, in units of eV cm$^{2}$, which contributes to multiple methods for sputtering. The quantity $\nu(E_{\textup{total}})$, divided by the number of nuclear collisions per incident ion, is the average energy lost per nuclear collision, or $W_{\textup{n}}$ \citep{johnson_energetic_1990}. $W_{\textup{n}}$ can be estimated by multiplying the energy of displacement, or the energy required to remove a species from the energy well of the surface, $E_{\textup{dis}}$, by 2.5 \citep{sigmund_theory_1969}. Depending on the path of the cascade collisions, and how deep the ice particle in question is located within the bulk of the ice, $E_{\textup{dis}}$ can range from the desorption energy, $E_{\textup{b}}$, for surface species,to 5$E_{\textup{b}}$ for species embedded in the crystal structure. We approximate $E_{\textup{dis}}$ for our physisorbed ice as $2E_{\textup{b}}$. This results in a $W_{\textup{n}}$ for water ice of approximately 2.45 eV, from an $E_{\textup{b}}$ of 0.49 eV \citep{garrod_complex_2008}. 

\begin{figure}
    \centering
    \includegraphics[width=\columnwidth]{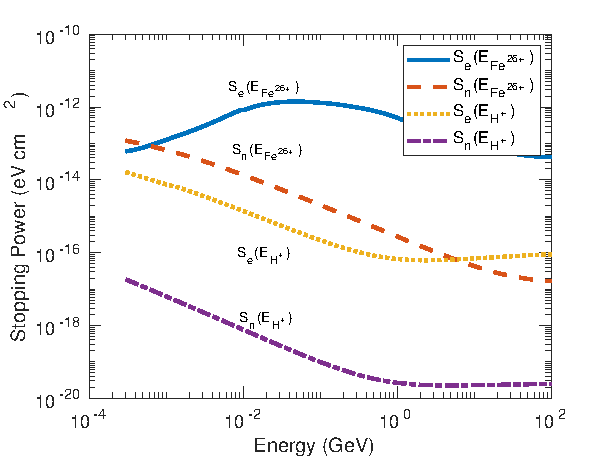}
    \caption{Electronic and nuclear stopping powers of H$^{+}$ and Fe$^{26+}$ ions impacting liquid water. H$^{+}$ electronic stopping powers are yellow dots, nuclear are purple dot-dashed. Iron incident ion electronic stopping powers are solid blue, nuclear are orange dashed. Energy ranges from $3 \times 10^{-4}$ GeV to 100 GeV. At these energies, $S_{e}$ of hydrogen is consistently $>10^{2}$ than the $S_{n}$. Liquid water is used as an approximation for amorphous solid water, which is not included in the \texttt{SR-NIEL} web tool. Calculated using \texttt{SR-NIEL} web tools \protect\hyperlink{http://www.sr-niel.org/index.php/}{http://www.sr-niel.org/index.php/}. }
    \label{fig:stopping_powers}
\end{figure}

The cross sections $\sigma_{n}$ and  $\sigma_{e}$ are approximated from the nuclear and electronic stopping power cross sections $S_{\textup{n}}$, and $S_{\textup{e}}$. Figure ~\ref{fig:stopping_powers} shows a graph of nuclear and electronic stopping powers for hydrogen and iron ions impacting water ice, as approximated using the values for liquid water and calculated by the \texttt{SR-NIEL} web tool \citep{boschini_screened_2014}. Stopping powers are dependent on the mass and energy of the incident ion and the target material. Figure~\ref{fig:stopping_powers} shows a three order of magnitude difference between electronic stopping powers and nuclear stopping powers for incident hydrogen ions, and an increasing difference in stopping power with energy for incident iron ions. This suggests that the majority of energy deposited into grain ices by cosmic rays comes from electronic interactions, compared with nuclear interactions.  Figure~\ref{fig:stopping_powers} also shows that the stopping power of iron ions impacting water is significantly higher than hydrogen for both nuclear and electronic collisions. We will not further examine iron cosmic rays in this work however, as they are significantly less abundant compared with hydrogen cosmic rays \citep{cummings_galactic_2016, blasi_origin_2013}. In order to account for the differences in terms and physical causation for sputtering, we will continue to examine nuclear and electronic sputtering separately.

\subsubsection{Nuclear (Elastic) Sputtering}

Nuclear sputtering is caused via collisions, most notably in the keV region in which elastic collisions occur within the ice leading to rebounds and eventually sputtering \citep{johnson_energetic_1990}. This energy range is below that of cosmic rays that bombard dust particles. Throughout the cosmic ray energy distribution used in the paper ($3 \times 10^{-4}$ GeV to 100 GeV), nuclear stopping powers are approximately three orders of magnitude lower than electronic stopping powers for water ice. For use in our models, which do not account for differing cosmic ray ions and energies, we calculate a weighted average stopping power ($\bar{S}$) by the formula

\begin{equation}
    \label{eq:average_s}
    \bar{S} = \frac{\sum S(E)j(E)}{\sum j(E)},
\end{equation}

\noindent where $S(E)$ is the range of stopping powers over a range of energies (separated into nuclear and elastic stopping powers), and $j(E)$ is the Spitzer-Tomasko cosmic ray energy function. Our calculation results in an average nuclear stopping power for H$^{+}$ impacting water ice of $3.489 \times 10^{-18}$ eV cm$^{2}$. To convert our nuclear stopping power to a nuclear cross section, we simply divide $S_{\textup{n}}$ by $W_{\textup{n}}$, resulting in an average nuclear sputtering cross section of $1.424 \times 10^{-18}$ cm$^{2}$. The rate coefficient for nuclear sputtering can now be expressed as:

\begin{equation}
    \label{nuc_no_y}
    k_{ns} = \sigma_{n} \phi_{ISM} \approx \frac{S_{n}}{W_{n}} \phi_{ST} \frac{\zeta}{1.36 \times 10^{-17}}.
\end{equation}

\noindent which contains the assumption that only one particle will be ejected per incident ion, which may not be the case. A suitable yield term has been derived by \citet{sigmund_theory_1969} and modified by \citet{johnson_energetic_1990} resulting in the following expression for the nuclear scattering yield $Y_{\rm ns}$:

\begin{equation}
    Y_{ns} \approx \frac{3 \alpha S_{n}}{2 \pi^{2} \Bar{\sigma}_{\textup{diff}} U} 
    \label{eq:nuclear_yield}.
\end{equation}

\noindent This is an equation for cascade sputtering of a planar surface, where $\alpha$ is an experimentally determined fraction of the nuclear stopping power that is involved in cascade collisions that are close to the surface, $S_{\textup{n}}$ is the nuclear stopping power, $\Bar{\sigma_{\textup{diff}}}$ is the average diffusion cross section, and $U$ is the average binding energy in eV. For water ices being impacted by H$^+$ ions, Equation~(\ref{eq:nuclear_yield}) yields an average result of of $1.8 \times 10^{-3}$ particles per cosmic ray proton, with the parameters shown in Table~\ref{tab:sput_params}.

\begin{table}
    \centering
    \caption{Parameters for nuclear sputtering calculations on amorphous solid water.}
    \begin{tabular}{lcr}
    \hline
     Parameter & Value & Reference \\
    \hline
     S$_{n}$ (eV cm$^{2}$) & $3.489 \times 10^{-18}$ & This work \\
     $W_{n}$ & 2.45 eV & This work, \citep{sigmund_theory_1969} \\
     $\Bar{\sigma}_{\textup{diff}}$ (cm$^{2}$) & $~3.6 \times 10^{-16}$ & \citep{johnson_energetic_1990} \\
     U (eV) & 0.49 & \citep{garrod_formation_2006} \\
     $\alpha$ & 0.6 & \citep{andersen_sputtering_1981} \\
     Y$_{ns}$ & $1.8 \times 10^{-3}$ H$_{2}$O per H$^{+}$ & This work \\
    \hline
     $\sigma_{n}$ (cm$^{2}$) & $1.424 \times 10^{-18} $ & This work \\
     \hline
    \end{tabular}
    \label{tab:sput_params}
\end{table}

The final form of the rate coefficient for nuclear sputtering is 

\begin{equation}
    \label{nuclear_sputtering_final}
    k_{ns} = Y_{ns} \sigma_{n} \phi_{ISM},
\end{equation}

\noindent which, fully expanded, is as follows:

\begin{equation}
    \label{nuclear_sputtering_final_expanded}
    k_{ns}(s^{-1}) = \left( \frac{3 \alpha S_{n}}{2 \pi^{2} \Bar{\sigma}_{\textup{diff}} U} \right) \left( \frac{S_{n}}{W_{n}} \right) \left( \phi_{ST} \frac{\zeta}{1.36 \times 10^{-17}} \right).
\end{equation}

\subsubsection{Electronic (Inelastic) Sputtering}

Our methodology for determining the electronic sputtering rate coefficient ($k_\textup{{es}}$) is very similar to that for nuclear sputtering, with some nuclear terms replaced by electronic terms. We start with estimating electronic cross sections from electronic stopping powers. Similarly to the nuclear stopping power, we find an average electronic stopping power, weighted by the same cosmic ray energy distribution as nuclear sputtering, shown in Equation~(\ref{eq:average_s}). This results in an average electronic stopping power of $1.131 \times 10^{-15}$ eV per particle for the bombardment of water ice by H$^{+}$. 

To convert electronic stopping powers to electronic cross sections, we simply divide the average stopping power by the average energy lost to the surface by ion-pair generation ($W_{\textup{e}}$) and by the fraction of electronic energy lost through repulsive decay ($f_{\textup{e}}$) \citep{brown_erosion_1982,schou_transport_1980}. As far as we know, there is no comprehensive table of $f_{\textup{e}}$ terms, and these must be either estimated, or calculated for individual species. The product of $W_{\textup{e}} \times {f_{\textup{e}}}$ is also referred to as the average energy lost to thermal relaxation, or $\overline{\Delta E}$ (in eV) \citep{rook_electronic_1985}. The process results in an electronic cross section of $2.09 \times 10^{-16}$ cm$^{2}$, using a $W_{\textup{e}}$ of 27 eV \citet{shingledecker_general_2018} and an ${f_{\textup{e}}}$ of 0.2 from \citet{johnson_electronic_1982}. All of these values are for amorphous water ice.

$W_{\textup{e}}$ is well known for a variety of species \citep{fueki_reactions_1963} and can also be used to estimate cross sections from stopping powers for radiolysis reactions, as outlined in \citet{shingledecker_general_2018}. Related to $W_{\textup{e}}$, $W_{\textup{s}}$, the average energy lost to sub-excitation electrons, may be a good estimate for $\overline{\Delta E}$, the average energy lost to thermal excitation, assuming most of the energy of the sub-excitation electrons is lost to the surface as heat. Therefore, if $f_{e}$ is unknown, we can estimate the inefficiencies in energy transfer by using $W_{\textup{s}}$. The terms are organized in Table~\ref{tab:electronic_yield_terms}, along with parameters needed for calculating electronic yields (Y$_{\textup{es}}$),via the equation \citep{johnson_energetic_1990}

\begin{table}
    \centering
    \caption{Parameters for calculating electronic sputtering terms on amorphous solid water.}
    \begin{tabular}{lcr}
    \hline
    Parameter & Value & Reference \\
    \hline
        $\overline{S}_{e}$ (eV cm$^{2}$) & $1.131 \times 10^{-15}$ & This work \\
        W$_{e}$ (eV) & 27 & \citep{shingledecker_general_2018} \\
        $f_{e}$ (Unitless) & 0.20 & \citep{johnson_electronic_1982} \\
        $\sigma_{es}$ (cm$^{2}$) & $2.09 \times 10^{-16}$ & This work \\
        $C_{e}\times (f_{e}^{2})$ (Unitless) & $8 \times 10^{-4}$ & \citep{brown_linear_1980} \\
        n$_{B}$ (cm$^{-3}$) & $3.3 \times 10^{22}$& \citep{brown_linear_1980} \\ 
        U (eV) & 0.49 & \citet{garrod_formation_2006} \\
        $Y_{es}$ & 0.0045 H$_{2}$O per H$^{+}$ & This work \\
    \hline
    \end{tabular}
    \label{tab:electronic_yield_terms}
\end{table}

\begin{equation}
    \label{eq:electric_yield}
    Y_{es} \approx C_{e}\times (f_{e}^{2})[(n_{B}^{-1/3})(n_{B} \times \bar{S_{e}})/U]^{2}.
\end{equation}

\noindent Here $C_{e}$ is an experimental unitless constant, while $n_{B}$ (particles cm$^{-3}$) is used to approximate the average thickness of a monolayer of a surface species. The determined electronic sputtering yield is 0.0045 H$_{2}$O molecules per incident H$^+$ ion.

The final form of the rate coefficient for electronic sputtering is

\begin{equation}
    \label{eq:final_electronic_sputtering}
    k_{es} = Y_{es} \sigma_{es} \phi_{ISM},
\end{equation}

\noindent which, when fully expanded, takes the form:

\begin{equation}
    \label{final_electronic_exploded}
    k_{es} (s^{-1}) = \left( C_{e}\times f_{e}^{2} \left[ \frac{(n_{B}^{-1/3})(n_{B} \times \bar{S_{e}})}{U} \right]^{2} \right) \left( \frac{\overline{S}_{e}}{W_{e} f_{e}} \right) \left(  \frac{\phi_{st} \zeta}{1.36 \times 10^{-17}} \right)
\end{equation}

\subsubsection{Carbon Dioxide and Mixed Ice Sputtering}

\begin{table}
    \centering
    \caption{Parameters for CO$_{2}$ sputtering cross section and yield calculations}
    \begin{tabular}{lcr}
    \hline
     Parameter & Value & Reference \\
    \hline
     $\overline{S}_{e}$ (cm$^{2}$ eV) & $2.329 \times 10^{-15}$ & This work \\
     $W_{e}$ (eV) & 34.2 & \citep{johnson_electronic_1982} \\
     $f_{e}$ (Unitless) & 0.19 & \citep{johnson_electronic_1982} \\
     $\sigma_{\textup{es}}$ (cm$^{2}$) & $3.58 \times 10^{-16}$ & This work \\
     $C_{e}\times f_{e}^{2}$ (Unitless) & $1 \times 10^{-3}$ & \citep{brown_erosion_1982} \\
     $n_{\rm B}$ (cm$^{-3}$) & $2.3 \times 10^{-22}$ & \citep{johnson_energetic_1990} \\
     U (eV) & 0.22 & \citep{garrod_formation_2006} \\
     $Y_{es}$ & $0.073$ CO$_{2}$ per H$^{+}$ & This work \\
     \hline
    \end{tabular}
    \label{tab:co2_sput_params}
\end{table}

For calculating the sputtering parameters for H$^{+}$ impacting CO$_{2}$ ice, chosen as a common ice constituent that isn't water, we will only calculate the electronic sputtering terms, as nuclear sputtering is significantly less effective, as shown in the earlier calculations for water ice. The process is the same for water ice, outlined in the previous section, with constants and important terms used displayed in Table~\ref{tab:co2_sput_params}. Using Equation \ref{eq:average_s} for the range of $S(E)$ CO$_{2}$ values results in an average electronic carbon dioxide stopping power of $2.329 \times 10^{-15}$ eV cm$^{2}$. Dividing $\overline{S}_{e}$ by a $W_{e}$ of 34.2 eV and the $f_{e}$ of 0.19 \citep{johnson_electronic_1982}, we obtain the electronic sputtering cross section for CO$_{2}$ of $3.58 \times 10^{-16}$ cm$^{2}$. Applying CO$_{2}$ values in Table~\ref{tab:co2_sput_params} to Equation~\ref{eq:electric_yield} returns a sputtering yield of 0.073 carbon dioxide molecules per incident hydrogen ion. The higher yield for carbon dioxide ice compared to water ice is due to carbon dioxide ices' cross section being greater than water ice, and a significantly lower binding energy of carbon dioxide ice.

Our mixed ice model for sputtering simply adjusts the rate coefficient for sputtering based on the fractional amounts of water and CO$_{2}$ ice within the model, which leads to the following formula for any species of ice, N$_{i}$: 

\begin{multline}
    \label{eq:mixed_ice_rate}
    k_{es}^{\textup{mix}} = \left( k_{es(\textup{H}_{2}\textup{O})} \frac{N_{\textup{H}_{2}\textup{O}}}{N_{\textup{H}_{2}\textup{O}}+N_{\textup{CO}_{2}}} \right) N_{i} \\
    + \left( k_{es(\textup{CO}_{2})} \frac{N_{\textup{CO}_{2}}}{N_{\textup{H}_{2}\textup{O}}+N_{\textup{CO}_{2}}} \right)N_{i}.
\end{multline}

\noindent This treatment of a mixed ice is simplistic, as we do not account for mixed ice yields, binding energies, or stopping powers, all of which will vary from a single species ice.


\section{Results and Analysis}
\label{sec:results}

\begin{figure}
    \centering
    \includegraphics[width=\columnwidth]{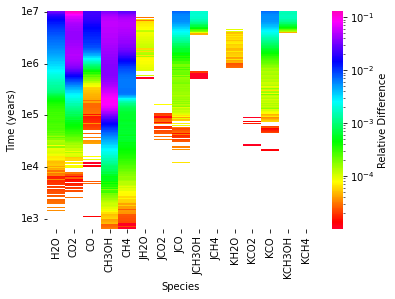}
    \caption{Heatmaps comparing models 1H and 2HS; each heatmap has model time on the y axis, with various species on the x axis. Species with a "J" designation represent surface ices, or a "K" for bulk ices. The different colours on the heatmap represent relative differences in abundance between models. Note the different colour bar scales and ranges}%
    \label{fig:sput}
\end{figure}

\begin{figure}
    \centering
    \includegraphics[width=\columnwidth]{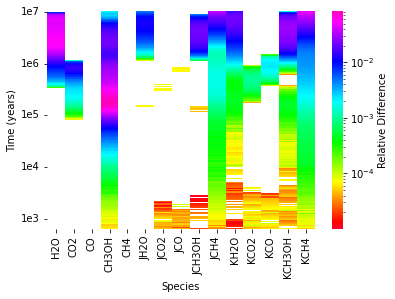}
    \caption{Heatmaps comparing models 1H and 3S; each heatmap has model time on the y axis, with various species on the x axis. Species with a "J" designation represent surface ices or a "K" for bulk ices. The different colours on the heatmap represent relative differences in abundance between species in the compared models. Note the different colour bar scales and ranges}%
    \label{fig:justsput}
\end{figure}

    

We utilize a series of low temperature models with various combinations of cosmic ray sputtering, cosmic ray whole grain heating, and cosmic ray induced diffusion described in Table~\ref{tab:model_key}. We model the physical conditions based on the cold molecular core TMC-1, where the initial abundances are compiled in Table~\ref{tab:abundances}, and relevant physical conditions are displayed in Table~\ref{tab:physicalcond}. The standard "reference" model is designated 1H, a basic model with the same chemical network as in \citet{paulive_role_2021}, without radiolysis chemistry. Models with an "H" after the number have cosmic ray whole grain heating, an "S" if sputtering is on, based on the theory in Section~\ref{sec:sputtering} for water ice, a "C" for sputtering based on CO$_{2}$ ice, and an "M" for the mixed ice models. The chemical networks for all models examined for these and all other models in this paper are the same.
All models that have CRH allow for periods of time where the temperature of the grain ices is temporarily increased to simulate the heating from cosmic rays, followed by cooling back to ambient temperatures as volatile species, such as CO, are thermally desorbed, as shown in \citet{hasegawa_new_1993}.
The rate coefficient for thermal desorption is the vibrational frequency of a species, $v_{\circ}$, multiplied by the negative exponent of the binding energy in kelvin, divided by the ambient dust temperature, resulting in a rate coefficient of

\begin{equation}
    k_{\textup{td}} = v_{\circ} \exp{ \left( -\frac{E_{\textup{b}}}{T_{\textup{dust}}} \right)}. 
\end{equation}

This is the same equation used to calculate the CRH rate coefficient, however, the ambient dust temperature is changed to the increased temperature used for CRH, in this case 70 K. This increased temperature results in an increase in desorption rate of  $e^{7}$ s$^{-1}$ compared to the thermal desorption rate at the dark cloud ambient dust temperature of approximately 10 K. 

\subsection{Water Ice Models}
\label{sec:water_results}

As an introduction, we examine the differences in the time-dependent results between the fiducial model (1H) with models 2HS and 3S. These non-fiducial models both include sputtering, while 2HS contains cosmic ray heating in addition to sputtering. 

Figures~\ref{fig:sput} and ~\ref{fig:justsput} show multiple heatmaps of a suite of common interstellar molecules found in TMC-1. The colour gradient on the heatmap shows the relative difference in molecular abundance between a species for a model with respect to the base model 1H, and the model to which it is being compared. The relative difference between two numbers X and Y can be defined by the ratio (Y-X)/X, where X applies to base model species and Y refers to other models such as 2HS and 3S. If Y is much larger than X, the relative difference approaches Y/X.

In Figure~\ref{fig:justsput}, the relative difference ranges from  $10^{-1}$ to $10^{-4}$, where a relative difference of $1$ represents a species with an abundance of Y that is double the abundance X in the base sputtering model. Likewise, a relative difference of $10^{3}$ would mean an abundance in model 2HS or 3S  1000 times the base model abundance, while a relative difference of $10^{-4}$ would be only signify an abundance 1.0001 times the abundance in model 1H. The white space within a heatmap signifies that there is no change in abundance between the two models at the given time.  

Specifically, for Figure~\ref{fig:sput}, these species are mostly abundant gas phase species within cold dark clouds , and we do not see differential large changes in their abundances, as they are generally efficiently produced in the gas phase, and while they could have efficient grain production pathways, the gas phase routes still dominate, or they are adequately desorbed from the grain surface with thermal and non-thermal methods. Looking at gas-phase CO, we see that there is an approximate 1\% increase toward the end of the model, as CO is often depleted onto grain surfaces at later times. With the addition of sputtering, we increase the amount of desorbed CO, resulting in slightly higher gas phase abundances. 



There are similar results for the comparison between 1H and 3S, shown in Figure~\ref{fig:justsput}, with minor differences: there is less of an increase in abundance in general, which is expected, because, unlike model 2HS, model 3S lacks cosmic ray whole grain heating. The lack of heating causes a lower desorption rate, which leads to the lower gas-phase abundance. In addition, the greater abundances in model 3S compared with 1H indicate that compared to whole grain heating already implemented in Nautilus, which temporarily heats the entirety of grain ices to 70 K, sputtering is generally more effective for desorbing low abundance species from the grain surface, and the bulk ice. This is due to sputtering allowing desorption directly from the bulk of the ice more readily than thermal evaporation, and thermal evaporation following CRH, as there is less of a chance that the ejected species can get trapped by the upper layers of ice from physical changes in the ice following ion bombardment.

Another contributing factor for the difference in abundances of models with solely CRH or sputtering is the relatively short time of increased thermal temperature from whole grain heating, which is on the order of $10^{-5}$ seconds, or the approximate time CO needs to thermally desorb from an ice surface \citep{hasegawa_new_1993}. During the duration of increased temperature, not all species are able to efficiently desorb at 70 K, but the more volatile CO is mostly desorbed from a surface in the time period of $10^{-5}$ seconds \citep{hasegawa_new_1993}. CO, and other species that are able to significantly contribute to desorption at 70, then cool the surface to the ambient 10 K. There is no equivalent duration or limitation on what species are effectively removed by sputtering, save for the cosmic ray-grain encounter rate.

\begin{figure}
    \centering
    \includegraphics[width=\columnwidth]{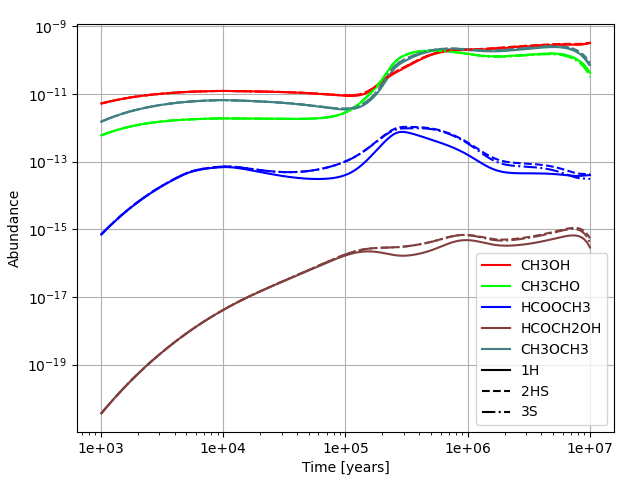}
    \caption{Modelled fractional abundance of relevant COMs under cold dark cloud conditions at 10 K.  Models 2HS and 3S have water ice sputtering rates of $2.55 \times 10^{-18}$ cm$^{-3}$ s$^{-1}$ at $5\times10^{5}$ yr. The dashed line is model 2HS, the solid line is model 1H, the dotted-dashed line is model 3S. All COMs examined for this graph are gas-phase. Both sputtering models show increases in abundances for methyl formate and glycoaldehyde.}
    \label{fig:water_sput_coms}
\end{figure}

In Figure ~\ref{fig:water_sput_coms} we show the modelled fractional abundance against time using the models 1H, 2HS, and 3S for a variety of COMs detected toward TMC-1. The models are represented by dashed lines, solid lines, and dotted dashed-lines for models 2HS, 1H, and 3S respectively. The molecules examined are methanol (CH$_{3}$OH), acetaldehyde (CH$_{3}$CHO), methyl formate (HCOOCH$_{3}$), glycolaldehyde (HCOCH$_{2}$OH), and dimethyl ether (CH$_{3}$OCH$_{3}$. All fractional abundances mentioned in this paper are with respect to H$_{2}$. The molecules, which will be discussed later, have binding energies that prevent them from being efficiently desorbed either at the nominal 10 K, or 70 K in models with CRH. 
There are little to no differences in the abundances of methanol, acetaldehyde, and dimethyl ether among models 1H, 2HS, and 3S at most times. However, there is an increase in abundance by more than a factor of 2, for methyl formate at $5 \times 10^{5}$ yr between model 1H and models 2HS and 3S. There is also an increase in gas-phase glycolaldehyde by less than a factor of 2 at the same time.

\subsection{Carbon Dioxide Ice Models}
\label{sec:co2_results}

\begin{figure}
    \centering
    \includegraphics[width=\linewidth]{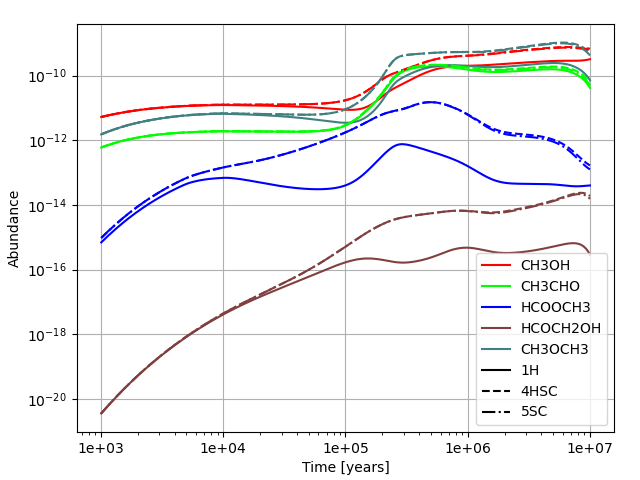}
    \caption{Gas-phase  fractional abundances of select COMs with respect to hydrogen. Models 4HSC and 5SC have carbon dioxide ice sputtering rates of $1.48 \times 10^{-16}$ cm$^{-3}$ s$^{-1}$ at $5\times10^{5}$ yr. Different colors represent different species. The fiducial model (1H) has a solid line, the model with carbon dioxide sputtering and heating (4HSC) a dashed line, and the model with just sputtering is has a dotted and dashed line (5SC). All molecules presented here show increases of varying degrees in gas-phase abundances, with acetaldehyde (CH$_{3}$CHO) showing less of an increase  (a factor of 1.12) compared with other COMs when comparing abundnances between the models with sputtering and the fiducial model.}
    \label{fig:co2_sput_coms}
\end{figure}

Table~\ref{tab:co2_sput_params} shows relevant parameters used in calculating yields based on carbon dioxide sputtering. We examine the same suite of molecules as in the previous section, shown in Figures~\ref{fig:co2_ices_sput} and ~\ref{fig:co2_sput_coms}. Figure~\ref{fig:co2_ices_sput} show a sample of common interstellar ices and simple molecules found within the ice. Notably, the addition of sputtering does not significantly alter the bulk or surface abundances of the species. This suggests that sputtering does not remove enough material to leave the grain barren of ice when coupled with previously examined thermal, chemical, and photo-desorption methods for these common species at 10 K. 

%

In contrast, Figure~\ref{fig:co2_sput_coms} shows significant increases in gas-phase abundances of the species presented in the figure for both models that include sputtering. Slight differences, especially at later times in the model, become apparent, with model 4HSC (Both sputtering and cosmic ray heating) slightly outproducing glycolaldehyde (HCOCH$_{2}$OH) and methyl formate (HCOOCH$_{3}$) compared to model 5SC (sputtering, no CRH).

Overall, we find significant differences between sputtering at carbon dioxide rates and water rates, partially due to higher yields and larger cross sections, leading to faster sputtering and greater amounts of desorption. Figure~\ref{fig:co2_sput_coms} shows increases of varying magnitude for COMs, where acetaldehyde has the lowest increase in magnitude by a factor of 1.12, while methyl formate and glycolaldehyde see increases by factors of 33 and 24, respectively. These increases contrast with our water sputtering models that only show increases in methyl formate and glycolaldehyde by an approximate factor of 2. For the other two examined COMs, methanol and dimethyl ether, we see increases by factors of 2.1 and 2.7, respectively, at the time $5\times10^{5}$ yr.



\subsection{Mixed Ice Models}
\label{sec:mixed_ice}

\begin{figure}
    \centering
    \includegraphics[width=\linewidth]{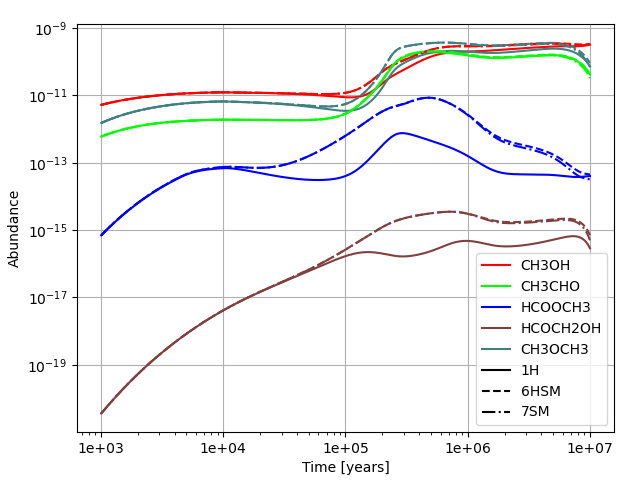}
    \caption{Gas-phase fractional abundances of select COMs with respect to hydrogen. Models 6HSM and 7SM have mixed water and carbon dioxide ice sputtering rates of $7.02 \times 10^{-17}$ cm$^{-3}$ s$^{-1}$ for water and $8.21 \times 10^{-17}$ cm$^{3}$ s$^{-1}$ for carbon dioxide at $5\times10^{5}$ yr. Different colors represent different species, the fiducial model (1H) is a solid line, the model with carbon dioxide sputtering and heating (6HSM) is dashed, the model with pure sputtering is dotted and dashed (7SM). All molecules presented here show increases of varying degrees in gas-phase abundances, with the acetaldehyde (CH$_{3}$CHO) showing less of an increase compared to other COMs.}
    \label{fig:mix_sput_coms}
\end{figure}

\begin{figure}
    \centering
    \includegraphics[width=\linewidth]{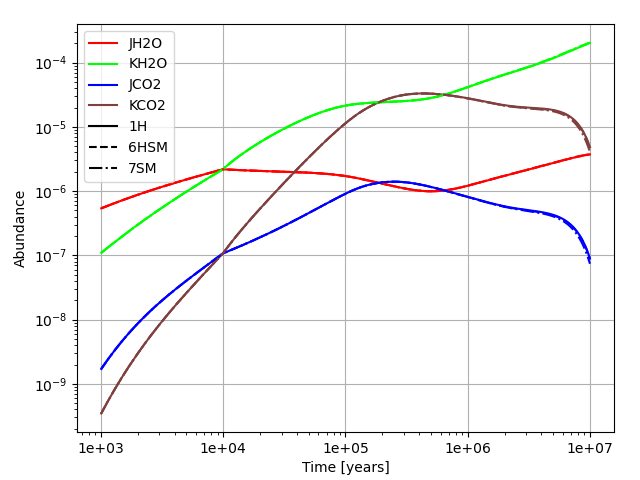}
    \caption{ Grain surface and mantle abundances of water ice and carbon dioxide ice, for models 1H, 6HSM, and 7SM. Other models used in this work differ in water and carbon dioxide fractional abundance by 10\% at most. The models from $1 \times 10^{3}$ yr and $1 \times 10^{6}$ yr have differences by factors of 0.0001, or approximately 0.01\%.}
    \label{fig:mixicecomp}
\end{figure}

The models containing the mixed ice sputtering rates (Equation~\ref{eq:mixed_ice_rate}) are labelled 6HSM and 7SM, where both include mixed ice sputtering rates while only 6HSM has whole grain cosmic ray heating. Figures~ \ref{fig:mix_sput_coms} and \ref{fig:mix_common_sput} shows similar curves to the carbon dioxide ice sputtering results, with slight variation in most common and simple ice species, albeit to a lesser extent than the carbon dioxide sputtering models. Similar to previous results, the mixed ice has desorption rates that vary based on the ratio of carbon dioxide to water ice which results in gas-phase abundances between those of pure water ice and pure carbon dioxide ice sputtering. Figure ~\ref{fig:mixicecomp} shows the difference in abundance of mantle and surface water ice and carbon dioxide ice in models 1H, 6HSM, and 7SM. The differences between ice abundances in the models is approximately 0.01\% for most times. There are no new notable results that have not been discussed in the two previous sections, apart from slightly different peak abundances for the examined COMs, shown in Figure~\ref{fig:mix_sput_coms}; these results will be further discussed in the next section, where we compared modelled abundances to observations.

\section{Astrochemical Implications}
\label{sec:analysis}

\begin{table*}
    \centering
    \caption{ List of observed column densities and estimated fractional abundances for examined COMs in TMC-1 methanol peak (MP) and cyanopolyyine peak (CP).  Column densities are in cm$^{-2}$, while fractional abundances are with respect to hydrogen, and unitless.}
    \begin{tabular*}{\textwidth}{lcccc}
    \hline 
    TMC-1 \\
    \hline 
     Molecule & Column Density (CP) & Fractional Abundance (CP) & Column Density (MP) & Fractional Abundance (MP) \\
    \hline 
    Methanol (CH$_{3}$OH) & $(1.7 \pm 0.3) \times 10^{13}$ & $(1.7  \pm 0.3) \times 10^{-9}$ & $(4.0 \pm 1.4) \times 10^{13}$ & $(4.0 \pm 1.4) \times 10^{-9}$ \\
    Acetaldehyde (CH$_{3}$CHO) & $(3.5 \pm 0.2) \times 10^{12}$ & $(3.5 \pm 0.2) \times 10^{-10}$ & $(3.4 \pm 0.4) \times 10^{12}$ & $(3.4 \pm 0.4) \times 10^{-10}$ \\
    Methyl Formate (HCOOCH$_{3}$) &  $(1.1 \pm 0.2) \times 10^{12}  $ & $(1.1 \pm 0.2) \times 10^{-10}$  & $(1.9 \pm 0.2) \times 10^{12}$ & $(1.9 \pm 0.2) \times 10^{-10}$ \\
    Dimethyl Ether (CH$_{3}$OCH$_{3}$) & $(2.5 \pm 0.7) \times 10^{12}$ & $(2.5 \pm 0.7)  \times 10^{-10}$ & $(2.1 \pm 0.7) \times 10^{12}$ & $(2.1 \pm 0.7) \times 10^{-10}$ \\
     \hline
     \end{tabular*}
     \\
     \begin{flushleft}Estimated hydrogen column density is $10^{22}$ cm$^{-2}$. All column densities are from \citet{soma_complex_2018}, except for methyl formate and dimethyl ether at the cyanopolyyine peak \citep{agundez_o-bearing_2021} and acetaldehyde at the cyanopolyyine peak \citep{cernicharo_discovery_2020}. \end{flushleft} 
     \label{tab:com_observation}
\end{table*}

\begin{table*}
    \centering
    \caption{List of modeled peak fractional abundances for select  COMs, distinguished by model.}
    \begin{tabular}{lcccccccc}
    \hline
        Molecule & 1H & 2HS & 3S & 4HSC & 5SC & 6HSM & 7SM & 8HSC10 \\
    \hline
        Methanol & $2.8 \times 10^{-10}$ & $3.0 \times 10^{-10}$ & $2.9 \times 10^{-10}$ & $7.6 \times 10^{-10}$ & $7.3 \times 10^{-10}$ & $3.5 \times 10^{-10}$ & $3.4 \times 10^{-10}$ & $7.8 \times 10^{-9}$ \\
        Acetaldehyde & $1.9 \times 10^{-10}$ &  $1.9 \times 10^{-10}$ &  $1.9 \times 10^{-10}$ & $2.1 \times 10^{-10}$ & $2.1 \times 10^{-10}$ & $2.0 \times 10^{-10}$ & $2.0 \times 10^{-10}$ & $7.1 \times 10^{-10}$ \\
        Methyl Formate & $7.5 \times 10^{-13}$ & $1.1 \times 10^{-12}$ & $9.5 \times 10^{-13}$ & $1.5 \times 10^{-11}$ & $1.5 \times 10^{-11}$ & $8.5 \times 10^{-12}$ & $8.5 \times 10^{-12}$ & $1.7 \times 10^{-10}$ \\
        Dimethyl Ether & $2.4 \times 10^{-10}$ & $2.7 \times 10^{-10}$ & $2.6 \times 10^{-10}$ & $1.0 \times 10^{-9}$ & $1.0 \times 10^{-9}$ & $3.7 \times 10^{-10}$ & $3.7 \times 10^{-10}$ & $2.1 \times 10^{-8}$ \\
    \hline
    \end{tabular}
    \label{tab:peak_abundances}
\end{table*}

Comparing the results of the various models to astronomical observations, we find mixed results: most models without sputtering do not produce enough gas-phase molecules, and while sputtering does adequately increase the gas-phase abundance to observed abundances in some cases, though the faster sputtering models are not enough to adequately reproduce abunandances for other molecules. \cite{soma_complex_2018} reported column densities of methanol, dimethyl ether, and acetaldehyde at the cyanopolyyne peak (referred to as TMC-1 CP) as approximately $1.7 \times 10^{13}$, $4.6 \times 10^{12}$ and $2.0 \times 10^{12}$ cm$^{-2}$, respectively. They also report methanol peak (TMC-1 MP) column densities of methanol, acetaldehyde, methyl formate, and dimethyl ether as $4.0 \times 10^{13}$, $3.4 \times 10^{12}$, $1.9 \times 10^{12}$, and $2.1 \times 10^{12}$ cm$^{-2}$, respectively. \citet{agundez_o-bearing_2021} provides column densities for methyl formate and dimethyl ether toward TMC-1 CP of $1.1 \times 10^{12}$ and $2.5 \times 10^{12}$ cm$^{-2}$, respectively, while \citet{cernicharo_discovery_2020} provides an updated column density for cyanopolyyine peak acetaldehyde of $3.5 \times 10^{12}$ cm$^{-2}$ Dividing these column densities by an approximate molecular hydrogen column density of $10^{22}$ cm$^{-2}$ in TMC-1 yields fractional abundances of the examined molecules. The observed values used here are organized in Table~\ref{tab:com_observation}, while peak modeled abundances are in Table~\ref{tab:peak_abundances}.

This results in an observed fractional abundance for methanol of  $1.7 \times 10^{-9}$ toward TMC-1 CP, and an abundance of $4.0 \times 10^{-9}$ at TMC-1 MP. Compared with the peak abundances of methanol in model 1H of approximately $2.8 \times 10^{-10}$, we significantly under-produce methanol in TMC-1 by an approximate order of magnitude, at least in the gas phase. Our sputtering models have a peak gaseous abundance of $3.0 \times 10^{-10}$ in model 2HS, and $7.6 \times 10^{-10}$ in model 4HSC, both at times from $5 \times 10^{5}$ until $7 \times 10^{6}$ yr. Our mixed ice model is not as effective as the carbon dioxide ice, with a peak abundance of $3.5 \times 10^{-10}$. These results agree with the generally accepted methanol production method of hydrogenation of carbon on grain surfaces, because the inclusion of sputtering desorbs methanol already made, yet stuck within the grain ices, to better match observed abundances \citep{watanabe_efficient_2002, fuchs_hydrogenation_2009}. The models suggest that faster sputtering is needed to reproduce the observed abundances in TMC-1 MP because the mixed ice and water ice sputtering rate models still under-produce methanol by factors of 11.5 and 13, respectively, when comparing peak abundances. In contrast, CO$_{2}$ sputtering seems to produce and desorb enough CH$_{3}$OH to match peak abundances in TMC-1 CP within a factor of 2.5 lower than the observed abundance, despite still under-producing methanol for TMC-1 MP by a factor of 5.3.

Acetaldehyde is an order of magnitude less abundant compared to methanol at both peaks, with peak observed abundances of $3.5 \times 10^{-10}$ at TMC-1 CP \citep{cernicharo_discovery_2020} , and $3.4 \times 10^{-10}$ toward the methanol peak. Our modelled abundances for acetaldehyde are $1.9 \times 10^{-10}$ for the standard heating model of 1H. For models including sputtering, the abundances are as follows: model 2HS has a peak abundance of $1.9 \times 10^{-10}$, model 4HSC peaks at $2.1 \times 10^{-10}$ and model 6HSM peaks at about $2.0 \times 10^{-10}$, with the faster carbon dioxide ice sputtering and heating better matching the observed abundances of the CP. While the base model is within a factor of 1.9 to the peak abundances toward TMC-1 CP, this model also under-produces such abundances for TMC-1 MP by a factor of 1.8; sputtering partially accounts for the difference for TMC-1 MP, but is less than the observed abundance by a factor of 1.6. The calculated peak abundances occur at approximately $5 \times 10^{5}$ yrs, with another peak of similar abundances occur at approximately $6 \times 10^{6}$ yrs in all models examined in this paper, and are shown in Figures ~\ref{fig:water_sput_coms}, ~\ref{fig:co2_sput_coms}, and ~\ref{fig:mix_sput_coms}.

The observations for methyl formate have the fractional abundance at $1.9 \times 10^{-10}$, toward the TMC-1 MP. Model 1H calculates the peak fractional abundance as $7.5 \times 10^{-13}$, significantly lower than the observed amount by over two orders of magnitude. The models with sputtering come closer to replicating the observed abundance, but still under-produce methyl formate in the gas phase by an approximate order of magnitude, with model 4HSC just reaching $1.5 \times 10^{-11}$. The mixed-ice and water models produce less methyl formate, producing peak amounts of $8.5 \times 10^{-12}$ and $1.1 \times 10^{-12}$, respectively. All models predict the peak of methyl formate to be at a similar time of $5 \times 10^{5}$ yrs, much like the other molecules highlighted in this section. Recent observations by \citet{agundez_o-bearing_2021} have detected methyl formate toward TMC-1 CP, with a column density of $ (1.1 \pm 0.2) \times 10^{12}$ cm$^{-2}$, and a fractional abundance of $1.1 \times 10^{-10}$. The results comparing the models to these observed amounts match the discrepancies between the models and the observed abundances in TMC-1 MP.
Methyl formate has also been examined in cold dark clouds using radiolysis models, such as in \citet{shingledecker_cosmic-ray-driven_2018}, where gas-phase methyl formate production was found to be greatly enhanced with the inclusion of radiolysis chemistry and suprathermal reactions. It may be possible to combine the enhanced grain ice abundances of ices in general, and efficiently desorb them using sputtering, as a future possibility.

Dimethyl ether is reported to have an upper limit to the fractional abundance of $4.6 \times 10^{-10}$ toward the cyanopolyyine peak, and an observed fractional abundance of $2.1 \times 10^{-10}$ toward the methanol peak \citep{soma_complex_2018}. In the same publication for methyl formate detection toward the cyanopolyyine peak, \citet{agundez_o-bearing_2021} report a column density of $(2.5 \pm 0.7) \times 10^{12}$ cm$^{-2}$, for a fractional abundance of $(2.5 \pm 0.7) \times 10^{-10}$. All models reach peak abundance at approximately $6 \times 10^{6}$ yrs. Model 4HSC produces the most dimethyl ether, reaching just above $1 \times 10^{-9}$, while 2HS and 6HSM have peak abundances at $2.7$ and $ 3.7 \times 10^{-10}$ respectively. However, at a time of $~5 \times 10^{5}$ yrs, which is similar to other peak times that match observations of molecules discussed earlier in the section, the modelled abundances are lower by a factor of approximately 2 than later in the model. 1H produces $~2.0 \times 10^{-10}$, with 4HSC producing approximately $5 \times 10^{-10}$ dimethyl ether. Models 2HS and 6HSM produce abundances of $~2.2 \times 10^{-10}$ and $~3.6 \times 10^{-10}$ respectively at 10$^5$ yrs. Models 4HSC and 5SC produce dimethyl ether within a factor of 4 toward both peaks of TMC-1, at peak abundances, with the other models all producing dimethyl ether within a factor of 1 at peak abundances. The carbon dioxide ice models more closely match observations at times around $5 \times 10^{5}$ years, within a factor of 2 for the cyanopolyyine peak, and more than 2 for the methanol peak.

Overall, the model that best fits the observed abundances for COMs in TMC-1 is either 4HSC or 5SC, despite under-producing methanol values for the MP by a factor of five. We came closer to replicating methanol values for the CP, only under-producing them by a factor of 2. For acetaldeyhde, all models match the observations for the CP, approximately $2 \times 10^{-10}$.For methyl formate at the MP, the carbon dioxide ice models under-produce methyl formate by a factor of 10, with other models under-producing methyl formate  by a larger factor. At TMC-1 CP, models 4HSC and 5SC also under-produce methyl formate by an order of magnitude. Finally, the upper limit of observed dimethyl ether toward the CP are less than peak amounts in models 4HSC and 5SC by a factor of 2. The peaks of other models are below  the upper limit of dimethyl ether observed in the CP, while slightly overproducing dimethyl ether compared with the observed abundance in the MP by a factor of 2. The models with sputtering rates based on carbon dioxide ices consistently under-produce the observed molecules examined in this paper, and their calculated abundances seem to best replicate the abundances found toward the TMC-1 CP.

These COMs can be compared with observed abundances for other environments in the ISM: the cold dense core B1-b, the dense core L483, the prestellar core L1689B, and the molecular cloud Barnard 5, as listed in Appendix~\ref{sec:appa}, Tables ~\ref{tab:B1b}, ~\ref{tab:L483}, ~\ref{tab:L1989B}, and ~\ref{tab:barnard5}. These environments have slightly different initial abundances and conditions compared with TMC-1, though the model parameters remain the same. Comparing modelled abundances with the observed abundances, we see that our variety of models does not adequately replicate methanol abundances in most environments. The observed methanol abundances range from $1.2 \times 10^{-9}$ in L1989B, to $4.5 \times 10^{-8}$ toward Barnard 5. However, the abundances of the other molecules examined are relatively well matched at peak, or near-peak abundances.

To compare methanol fractional abundances to the approximate observed levels in all sources would require the rate coefficient for sputtering to be increased by approximately a factor of ten. We have run a supplementary model (8HSC10) with such an increase, and have included the figures presenting the fractional abundances for ten times the sputtering rate in Appendix ~\ref{sec:appa}, Figure ~\ref{fig:go_crazy}. Notable in this model is that the abundance of dimethyl ether is greatly increased, to a greater fractional abundance than that of methanol, which is not observed in any of the sources mentioned previously. However, at this enhanced rate of sputtering, peak modelled methanol abundances match observed abundances well. This suggests that while sputtering rates greater than the ones presented here may be physical, there need to be further studies into modelling the production and destruction of the molecules highlighted here, especially methanol and dimethyl ether, and further examinations of sputtering.


\section{Conclusions}
\label{sec:conclusions}

We present in this paper a way of both estimating and implementing theoretical sputtering parameters into a rate-equation based model of cold dark clouds. We show that sputtering is an effective way of desorbing multiple species off of grain surfaces, even at slow sputtering rates. While there are experimental results that can be included in models, it is more difficult to obtain experimental data that match the mixture of amorphous ices that populate grain surfaces in the ISM. Fortunately, there are theoretical treatments of sputtering ices that seem to be reasonably effective. While further experimental work will need to be done to provide a basis for the sputtering of mixed ices by lighter ions, both current theory and experimental results suggest that sputtering is important as a method of non-thermal desorption in astrochemical models, similar to reactive desorption and photodesorption. Even when coupled with cosmic ray heating, a widely used "thermal" method of temporarily increasing thermal desorption, many species are not excessively depleted from within the ice, nor excessively ejected into the gas phase. In addition, in cases where there are efficient gas destruction pathways, sputtering does not overcome multiple efficient gas phase destruction routes. To adequately match abundances using sputtering alone, the rate coefficient for sputtering will need to be increased by a factor of approximately 10, however, this causes overproduction of multiple molecules including dimethyl ether and acetaldehyde. Further examination of gas and grain destruction pathways is warranted, should sputtering become commonplace in models.

\section*{Acknowledgements}

E. H. thanks the National Science Foundation (US) for support of his research programme in astrochemistry through grant AST 19-06489. This research has made use of NASA's Astrophysics Data System Bibliographic Services. We would like to thank V. Wakelam for the use of the \texttt{Nautilus-1.1} program.

\section*{Data Availability}
The data underlying this article are available in the article, as well as cited online repositories (KIDA). 




\bibliographystyle{mnras}
\bibliography{references,chris_references,sputtering_references}



\appendix
\section{Graphs and Tables}
\label{sec:appa}

Appendix A includes large figures and tables that are repetitive, but still contain relevant information, due to slight variations in models. The large figures show ice abundances of common species, with surface ices denoted with a "J", and bulk ices identified with a "K". We find that these ices are less affected by sputtering compared to other ices because of greater fractional abundances. The more abundant molecules have less of a noticeable increase in desorption compared to ices with a lower abundance. The tables contain fractional abundances and column densities for B1-b, L483, L1989b, and Barnard 5.

\begin{table}
    \centering
    \caption{List of observed column densities in cm$^{-2}$ for selected molecules toward B1-b, with a reported H$_{2}$ column density of ~$10^{23}$ cm$^{-2}$ \citep{cernicharo_discovery_2012}.}
    \begin{tabular}{lcc}
     \hline 
     Compound & Column Density & Fractional Abundance \\
     \hline 
     CH$_{3}$OH$^{\rm a}$ & $~3.6 \times 10^{14}$ & $3.608 \times 10^{-9}$ \\
     CH$_{3}$CHO$^{\rm a}$ & $5.4 \times 10^{12}$ & $3.9 \times 10^{-11}$ \\
     HCOOCH$_{3}^{\rm a}$ & $8.3 \times 10^{12}$ & $8.3 \times 10^{-11}$ \\
     CH$_{3}$OCH$_{3}^{\rm b}$ & $3.0 \times 10^{12}$ & $2.5 \times 10^{-10}$ \\
     \hline
      a \citep{oberg_cold_2010} \\
     b \citep{cernicharo_discovery_2012}
     \end{tabular}

     \label{tab:B1b}
\end{table}

\begin{table}
    \centering
    \caption{List of observed column densities in cm$^{-2}$ for selected molecules toward L483. All observed abundances are from \citet{agundez_sensitive_2019}, and with a H$_{2}$ column density of approximately $4 \times 10^{23}$ cm$^{-2}$ }
    \begin{tabular}{lcc}
     L483 \\
     \hline 
     Compound & Column Density & Fractional Abundance \\
     \hline 
     CH$_{3}$OH & $2.9 \times 10^{13}$ & $7.25 \times 10^{-9}$ \\
     CH$_{3}$CHO & $3.8 \times 10^{12}$ & $9.5 \times 10^{-11}$ \\
     HCOOCH$_{3}$ & $2.3 \times 10^{12}$ & $5.7 \times 10^{-11}$ \\
     CH$_{3}$OCH$_{3}$ & $5.3 \times 10^{12}$ & $1.32 \times 10^{-10}$ \\
     \hline 
       \end{tabular}
       \label{tab:L483}
\end{table}

\begin{table}
    \centering
    \caption{List of observed column densities in cm$^{-2}$ for selected molecules toward L1989B. The reported H$_{2}$ column density is $4 \times 10^{23}$ cm$^{-2}$.  Methanol detection is from \citet{bacmann_gas_2015} and other abundances are from \citet{bacmann_detection_2012}.}
    \begin{tabular}{lcc}
     L1989B \\
     \hline 
     Compound & Column Density & Fractional Abundance \\
     \hline  
     CH$_{3}$OH & $1.2 \times 10^{14}$ & $1.2 \times 10^{-9}$ \\
     CH$_{3}$CHO & $9.0 \times 10^{12}$ & $9.0 \times 10^{-11}$ \\
     HCOOCH$_{3}$ & $5.91 \times 10^{12}$ & $5.91 \times 10^{-11}$ \\
     CH$_{3}$OCH$_{3}$ & $4.6 \times 10^{12}$ & $5.91 \times 10^{-11}$ \\
     \hline 
       \end{tabular}
       \label{tab:L1989B}
\end{table}

\begin{table}
    \centering
    \caption{List of observed column densities in cm$^{-2}$ for selected molecules toward Barnard 5, H$_{2}$ column density of ~$10^{21}$ cm$^{-2}$ \citep{taquet_chemical_2017}. }
    \begin{tabular}{lcc}
     Barnard 5 \\
     \hline 
     Compound & Column Density & Fractional Abundance \\
     \hline 
     CH$_{3}$OH & $1.5 \times 10^{14}$ & $4.5 \times 10^{-8}$ \\
     CH$_{3}$CHO & $5.2 \times 10^{12}$ & $1.6 \times 10^{-9}$ \\
     HCOOCH$_{3}$ & $4.3 \times 10^{12}$ & $1.3 \times 10^{-9}$ \\
     CH$_{3}$OCH$_{3}$ & $2.1 \times 10^{12}$ & $7.0 \times 10^{-10}$ \\
     \hline 
     \end{tabular}
    \label{tab:barnard5}
\end{table}

\begin{figure*}
    \centering
    
    \subfloat[]{%
    \includegraphics[width=\columnwidth]{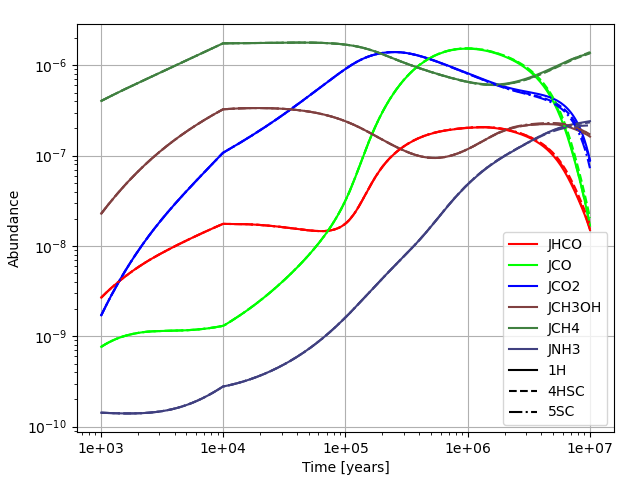}%
    \label{fig:co2_ices_sput:surf}%
    }\qquad
    \subfloat[]{%
    \includegraphics[width=\columnwidth]{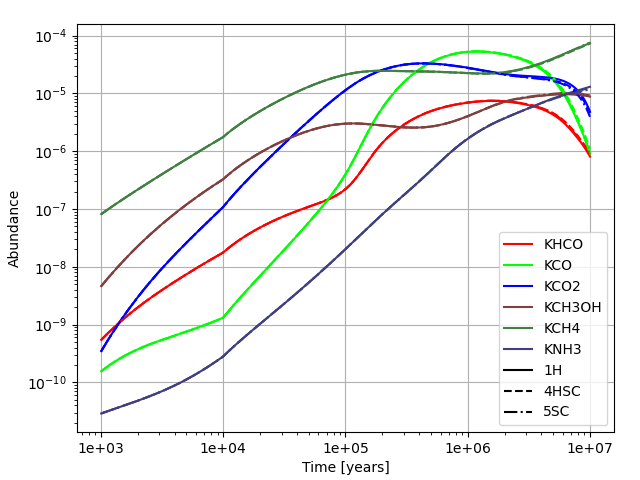}%
    \label{fig:co2_ices_sput:bulk}%
    }\qquad
    \subfloat[]{%
    \includegraphics[width=\columnwidth]{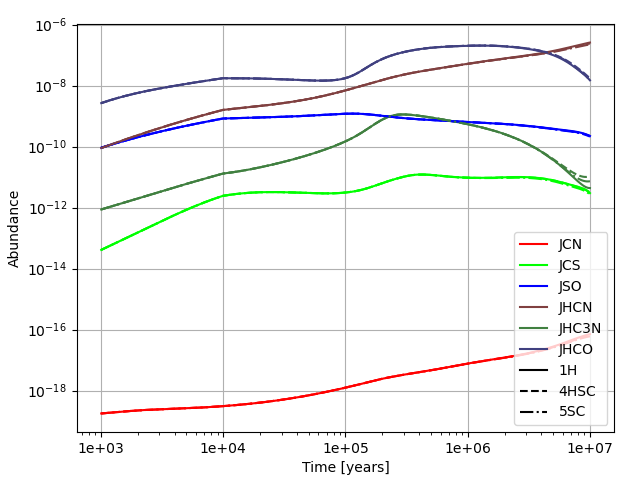}%
    \label{fig:co2_simp_sput:surf}%
    }\qquad
    \subfloat[]{%
    \includegraphics[width=\columnwidth]{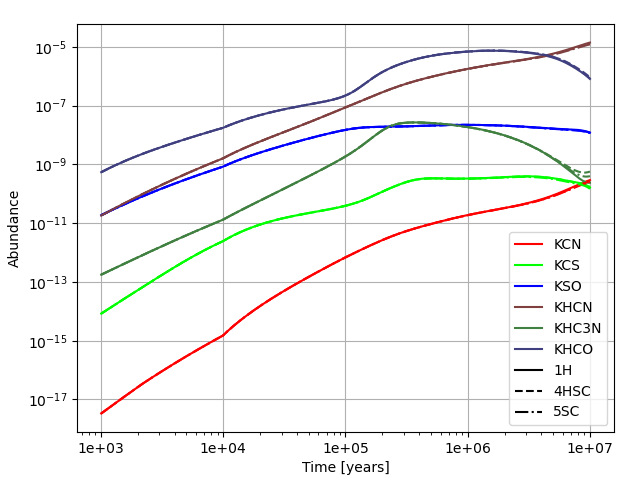}%
    \label{fig:co2_simp_sput:bulk}%
    }
    \caption{Fractional abundance graphs of various ice species with respect to hydrogen. At the right are the surface abundances of select species, at the left are the bulk abundances. Individual species are a specific colour.At the top are common ices, while the bottom shows simple ices. Model 1H, without sputtering, is a solid line, models 4HSC and 5SC are dashed and dotted-dashed lines, respectively. Notable in these figures is that the ice abundances are not significantly affected by sputtering, showing that sputtering is not fast enough to significantly deplete ices on the grain surface, even with an increased rate coefficient for sputtering of CO$_{2}$ ices.}%
    \label{fig:co2_ices_sput}
\end{figure*}

\begin{figure*}
    \centering
    
    \subfloat[]{%
    \includegraphics[width=\columnwidth]{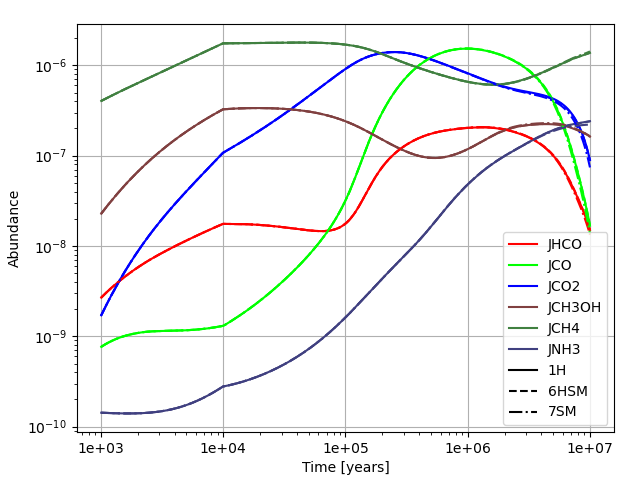}%
    \label{fig:mix_common_sput:surf}%
    }\qquad
    \subfloat[]{%
    \includegraphics[width=\columnwidth]{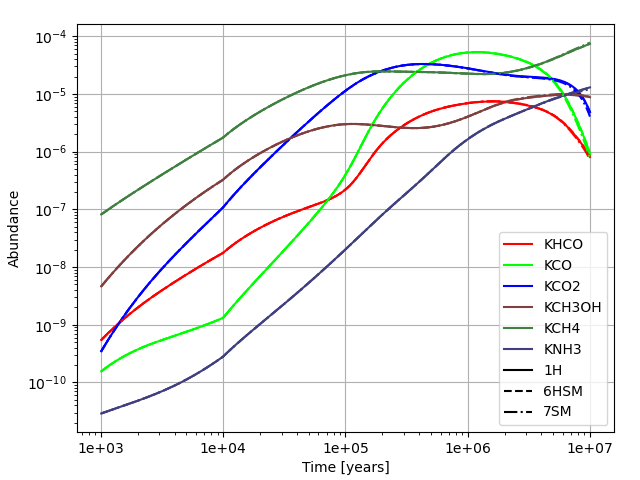}%
    \label{fig:mix_common_sput:bulk}%
    }\qquad
    \subfloat[]{%
    \includegraphics[width=\columnwidth]{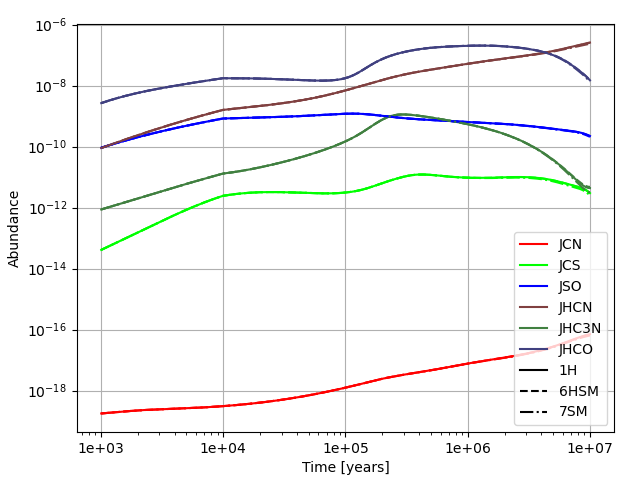}%
    \label{fig:mix_simp_sput:surf}%
    }\qquad
    \subfloat[]{%
    \includegraphics[width=\columnwidth]{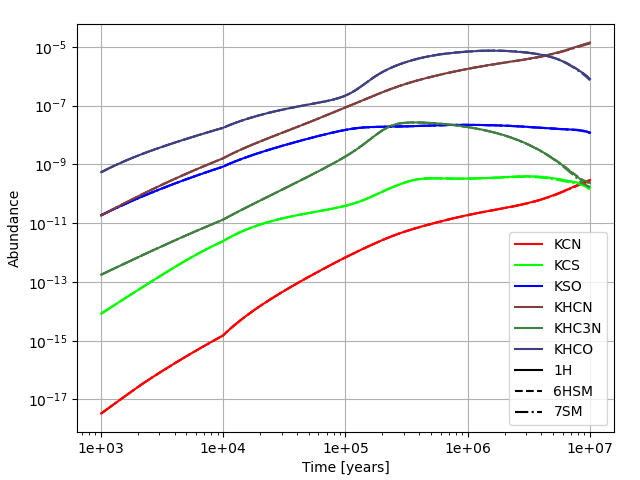}%
    \label{fig:mix_simp_sput:bulk}%
    }
    \caption{Fractional abundance graphs of various common ices with respect to hydrogen. At the right are the surface abundances of select species, at the left are the bulk abundances. Individual species are a specific colour. At the top are common ices, while the bottom is simple molecules. Model 1H, without sputtering, is a solid line, models 6HSM and 7SM are dashed and dotted-dashed lines, respectively. Notable in these figures is that the ice abundances are not significantly affected by sputtering, despite the increased sputtering rate coefficient of a mixed ice.}%
    \label{fig:mix_common_sput}
\end{figure*}

\begin{figure}
    \centering
    \includegraphics[width=\columnwidth]{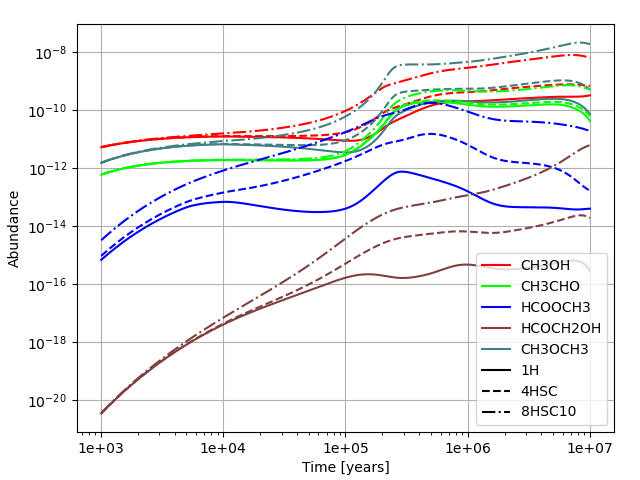}
    \caption{Plot of the fractional gas-phase abundance of various COMs with respect to hydrogen. Different species are different colours, the solid lines are model 1H, the dashed lines are model 4HSC, and the dotted-dashed lines are model 8HSC10, a model with 10 times the rate coefficient of sputtering in model 4HSC. The resulting rate of sputtering for carbon dioxide ices in model 8HSC10 is $1.42\times10^{-15}$ cm$^{-3}$ s$^{-1}$. }
    \label{fig:go_crazy}
\end{figure}


\bsp	
\label{lastpage}
\end{document}